\documentclass[graybox]{svmult}

\usepackage{amssymb}
\usepackage{amsmath}
\usepackage{cite}
\usepackage{balance}
\usepackage{booktabs}
\usepackage{float}
\usepackage{graphicx}
\usepackage{tikz,xcolor,hyperref}
\usepackage{subfigure}
\usepackage{multirow}
\usepackage{color}
\usepackage{float}
\usepackage{footnote}
\usepackage{stfloats}
\usepackage{array}
\usepackage{booktabs}
\usepackage[noend]{algpseudocode}
\usepackage{amsmath} 
\usepackage{algorithmicx,algorithm}
\usepackage{amsfonts} 
\usepackage{url}
\usepackage{balance}
\usepackage[misc]{ifsym} 
\usepackage{colortbl}  
\usepackage{xcolor}
\usepackage{array}   
\usepackage[justification=centering]{caption}
\usepackage{xcolor}
\newcommand{\hys}[1]{\textcolor{black}{#1}}

\usepackage{type1cm}        
\usepackage{makeidx}         
\usepackage{graphicx}        
\usepackage{multicol}        
\usepackage[bottom]{footmisc}

\usepackage{newtxtext}       %
\usepackage{newtxmath}       

\newcommand{\eg}{\textit{e.g.}}

\makeindex             

\begin{document}

\title*{ Infrared Image Super-Resolution via GAN}
\author{Yongsong HUANG and Shinichiro OMACHI}
\institute{Yongsong HUANG and Shinichiro OMACHI \at Graduate School of Engineering, Tohoku University, Sendai, Japan \\
\email{huang.yongsong.r1@dc.tohoku.ac.jp; machi@ecei.tohoku.ac.jp}}
%
%
\maketitle



\abstract{The ability of generative models to accurately fit data distributions has resulted in their widespread adoption and success in fields such as computer vision and natural language processing. In this chapter, we provide a brief overview of the application of generative models in the domain of infrared (IR) image super-resolution, including a discussion of the various challenges and adversarial training methods employed. We propose potential areas for further investigation and advancement in the application of generative models for IR image super-resolution.}

\section{Introduction\label{sec.1}}

In modern society, IR images play an irreplaceable role in industry and daily life. Although visible images make it easier to transfer information to people, in specific environments, such as earthquake rescue and security, where there is insufficient light, people have to turn to infrared images. Compared to visible images, infrared images can be better tolerated in tough natural environments and provide rich information about heat sources. Such valuable feedback can help people judge the condition from outdoor equipment and individuals in order to repair damaged infrastructure or help someone in distress. Considering these important applications, high-resolution (HR) IR images are needed urgently. However, IR image resolution is unsatisfactory due to the limitations by current optical devices. With the growing interest in Generative Adversarial Networks (GANs)\cite{goodfellow2020generative} in the deep learning community, introducing adversarial training methods for IR image super-resolution is on the agenda\cite{ledig2017photo,wang2018esrgan,ma2020structure,gulrajani2017improved}. In this section, we will present IR imaging applications first, and some key domain will be discussed. Then the basic components and challenges faced by infrared imaging systems are presented.

\subsection{Infrared Image Applications \label{sec.1.1}}

\begin{figure}[h]
\centerline{\includegraphics[width=\columnwidth]{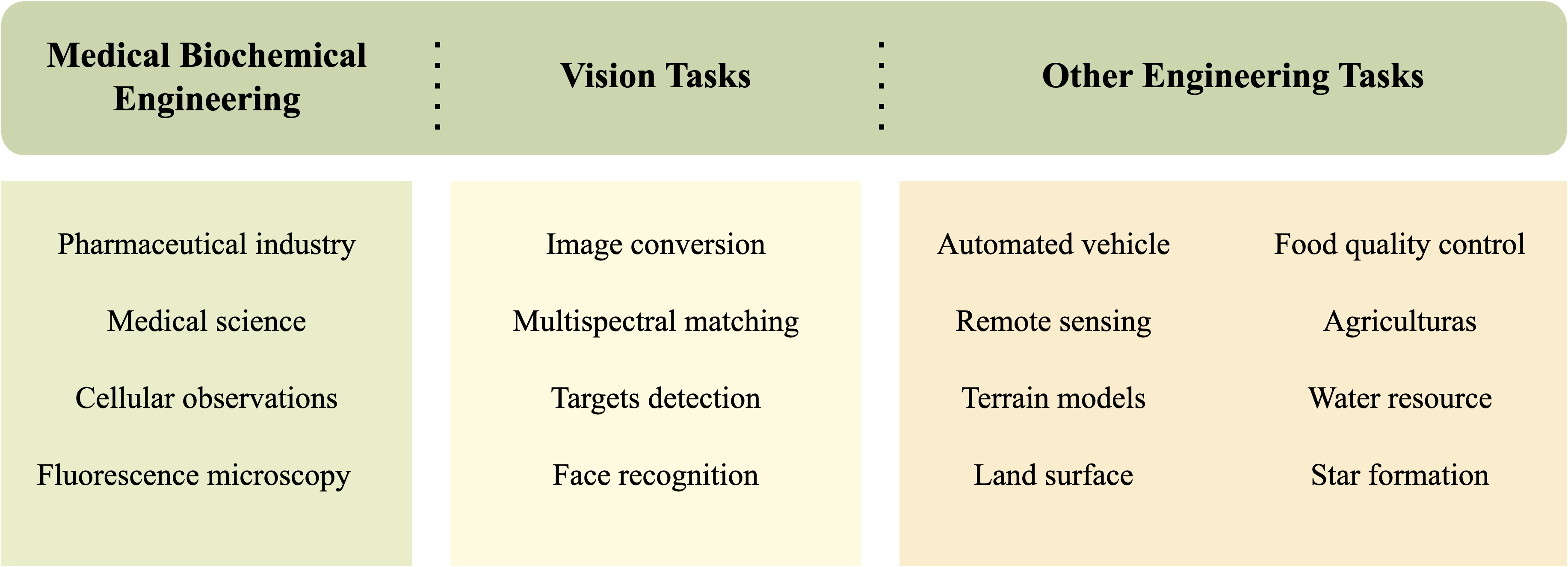}}
\caption{IR image super-resolution in different domains\cite{Infrared-Review}.}
\label{fig1}
\end{figure}

IR image super-resolution is an attractive approach in a wide range of realistic situations. We will present some typical fields of them, such as medical engineering and engineering tasks. Then, other methods will be briefly discussed. IR image super-resolution applications can be seen in Fig.\ref{fig1} with more details.

In the medical engineering field, degenerative diseases in the nervous system have been gaining great attention. If the mechanisms behind such diseases can be clarified, we will have the opportunity to completely treat diseases that significantly affect the older people's life quality in their later years, such as Alzheimer's disease. Before determining treatment options, understanding the mechanisms involved in these diseases is required for pathological analysis. One of the materials: CRANAD-2, the neurogenic curcumin derivative, is thought to contribute to these studies\cite{Torra2022VersatileNS,bouzin2022melanin}. IR images can detect CRANAD-2. High-resolution IR images would benefit for NIR nano imaging and further promote the development of correlational studies. Another representative example is the COVID-19 that has received much attention recently. Considering the imbalance in development between different countries and geographies, it will create an unbalanced medical resource. Further, impacting the early detection to COVID-19 in developing countries is that costly lung imaging equipment is not available. Many physicians have attempted to use faster and cheaper X-ray images as a tool for disease diagnosis\cite{Lukose2021OpticalTF,CanalesFiscal2021COVID19CU}.

IR imaging plays a significant role in the field of engineering. IR cameras can be used to detect temperature changes\cite{jang2022automated} in systems, which can help identify overheating components that may indicate potential failure or malfunction. In electrical engineering, this can be particularly useful in detecting overheating components that may cause power outages or other issues. IR imaging can also be used to detect leaks in systems that utilize gases or liquids, such as pipelines\cite{allred2021time}, allowing engineers to quickly locate and repair these issues and avoid costly damage and downtime. Additionally, IR imaging can be utilized to monitor the performance of systems and equipment, allowing engineers to identify potential problems before they become more serious and take action to prevent costly failures or downtime\cite{butkevich2021photoactivatable}.

Further, IR images have a crucial role in other fields as well. IR imaging can be used to monitor and evaluate the health of ecosystems\cite{lloyd2021optically,ping2021can,yang2020water} in the field of environmental protection, particularly in areas where it is difficult to access or study using traditional methods. IR cameras can provide useful information about the health of plants, animals, and other organisms by detecting temperature changes. This enables conservationists and scientists to identify potential difficulties and take measures to protect the environment. And, IR imaging can be used to monitor crops' health and productivity in the agricultural sector\cite{cao2020monitoring,liu2021simulating,qi2020estimation,barzin2021comparison}. IR cameras can help farmers identify areas of their fields that are not performing as well as others by detecting temperature changes. This enables farmers to take action to increase yields and lower the likelihood of crop failure. Additionally, identify diseases and pests, IR images enabling farmers to safeguard their crops and boost overall productivity. In the food industry,  IR cameras can also help identify areas of food that are not being properly cooked or stored by detecting temperature changes. This enables food manufacturers to take action to stop the spread of foodborne illness. As food products are being processed, IR imaging used to monitor their quality\cite{martinez2022advantage,wang2021identification}, allowing manufacturers to spot potential issues and improve product quality. In deep space exploration, IR cameras can provide useful information about the composition and structure of planets, moons, and other objects in the solar system by detecting temperature changes. This helps scientists better comprehend the universe's origins and evolution\cite{Biller2005HighResolutionMI,Megeath1996EvidenceFO}.

Overall, IR imaging is a versatile and powerful tool that can be used in a variety of different fields to provide valuable information and improve our understanding of the world around us.

\subsection{Fundamentals and Challenges in Infrared Imaging\label{sec.1.2}}

The link between temperature and the amount of IR radiation emitted by an item is one of the cornerstones of IR imaging\cite{rabal2018dynamic,lee2007image,huang2023target,jiang2021difference}. Objects generate more IR radiation at higher temperatures than they do at lower temperatures. Due to this connection, IR cameras may produce pictures based on the amount of IR radiation produced by objects in the field of vision and detect changes in temperature. The requirement for specialist equipment is one of the significant challenges in IR imaging. Special cameras or sensors are needed to find and quantify IR radiation because it is not visible to the naked eye. These cameras and sensors can be expensive, and their efficient use could need specific training. The requirement for meticulous instrument calibration and adjustment is another difficulty in IR imaging. IR cameras and sensors must be properly calibrated to guarantee that they are delivering accurate and dependable data, since they are sensitive to variations in temperature\cite{kim2018averaging,wang2016stripe,chen2014super,kong2013near,schutte2003signal}. This can take a lot of time and requires specialist knowledge. Additionally, a variety of variables, including air conditions, ambient temperature, and the existence of other IR radiation sources, can have an impact on IR imaging. When analyzing and utilizing IR pictures, these aspects must be carefully taken into consideration, since they have the potential to alter the images' accuracy and dependability.

\section{Adversarial Training Methods in Super-Resolution\label{sec.2}}

In this section, we will discuss adversarial training methods in super-resolution, including problem definition and adversarial training framework. For problem definition, we will introduce the components in super-resolution. Then, the adversarial training framework in super-resolution will also be shown.

\subsection{Problem Definitions\label{sec.2.1}}

\begin{figure}[htbp]
\centerline{\includegraphics[width=\columnwidth]{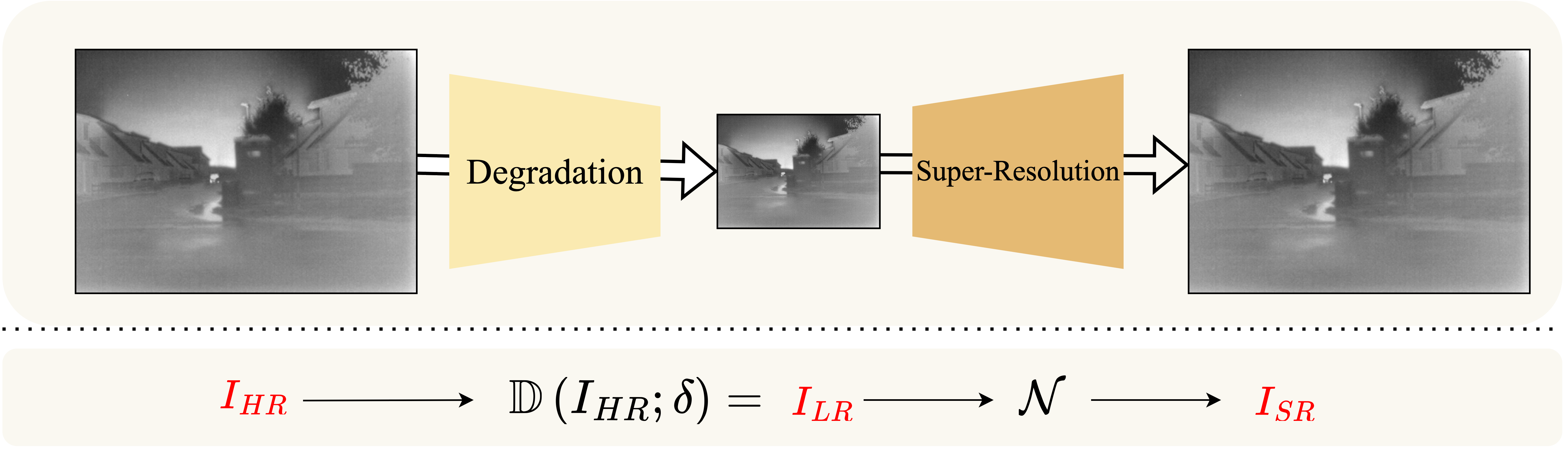}}
\caption{Degradation and reconstruction in IR image super-resolution. $\mathcal{N}$ denotes neural network.}
\label{fig2}
\end{figure}

Super-resolution, a technique in the fields of computer vision and image processing, involves the construction of high-resolution images from one or more low-resolution counterparts\cite{wang2020deep,park2003super,chen2022real,liu2022blind}. This method can be utilized in a variety of contexts, including the enhancement of images from low-resolution cameras or sensors and the augmentation of photographs for medical or scientific study. More
details of the problem definition are shown in Eq. \ref{eq.1}:

\begin{equation}
I_{L R}=\mathbb{D}\left(I_{H R} ; \delta\right)
\label{eq.1}
\end{equation}

where $\mathbb{D}$ denotes a degradation function, $I_{HR}$ is the high-resolution IR image, $I_{LR}$ is the low-resolution IR image and $\delta$ is the parameters of the degradation process. Visualization for the IR image super-resolution degradation and reconstruction can be found in Fig.\ref{fig2}.

\begin{equation}
    \mathbb{D}(I_{HR};\delta ) = (I_{HR}\otimes {k} )\downarrow _{d}; [{k,\downarrow _d}]\subset{\delta},
    \label{eq.2}
\end{equation}

where $I_{HR} \otimes {k} $ represents the convolution between a blur kernel ${k}$ and the $\mathrm{HR}$ image $I_{HR}$. In the ${k}$, noise and compression are included. And, $\downarrow _{d}$  is a downsampling factor, \eg, $4 \times$ and $8 \times$. Briefly, the super-resolution reconstruction objective function of IR images can be described as Eq. \ref{eq.3}:

\begin{equation}
\hat{\theta}=\underset{\theta}{\arg \min } \mathcal{L}\left(I_{HR}, I_{SR}\right)+\lambda \Phi(\theta),
\label{eq.3}
\end{equation}
where $\mathcal{L}$ denotes the loss function, between the HR image $I_{HR}$ and the SR image $I_{SR}$. $\Phi(\theta)$ and $\lambda$ are the regularization term and punishment parameter, respectively. More details about the IR image super-resolution definition can be found in this literature\cite{Infrared-Review}.

\subsection{Adversarial Training Framework\label{sec.2.2}}

The issue of super-resolution, which entails raising an image or video's resolution, has been addressed using GANs\cite{goodfellow2020generative,ledig2017photo,wang2018esrgan,ma2020structure,gulrajani2017improved}. In a super-resolution GAN, the discriminative model $D$ is trained to tell the difference between existing high-resolution instances $I_{HR}$ and produced ones $I_{SR}$, while the generative model $G$ is taught to create a high-resolution version of pictures or videos $I_{SR}$. A high-quality super-resolution effect is produced by the GAN when generated instances are indistinguishable from actual ones. The two models are trained in tandem using an adversarial process. The generative model ultimately improves greatly at providing realistic examples of high quality thanks to this back and forth training procedure. For super-resolution generative adversarial network (SRGAN), the objective function is shown in Eq.\ref{eq.4}. A feed-forward CNN $G$ parametrized by $\theta_G$ \hys{and} discriminator network $D$ which was defined by $\theta_D$, \hys{aim} to solve the adversarial min-max problem:

\begin{equation}
\begin{aligned}
\min _{\theta_G} \max _{\theta_D} & \mathbb{E}_{I_{HR} \sim p_{\text {train}}\left(I_{H R}\right)}\left[\log D_{\theta_D}\left(I_{H R}\right)\right]+ \\
& \mathbb{E}_{I_{L R} \sim p_G\left(I_{LR}\right)}\left[\log \left(1-D_{\theta_D}\left(G_{\theta_G}\left(I_{LR}\right)\right)\right]\right.
\end{aligned}
\label{eq.4}
\end{equation}

\begin{figure}[h]
\centerline{\includegraphics[width=\columnwidth]{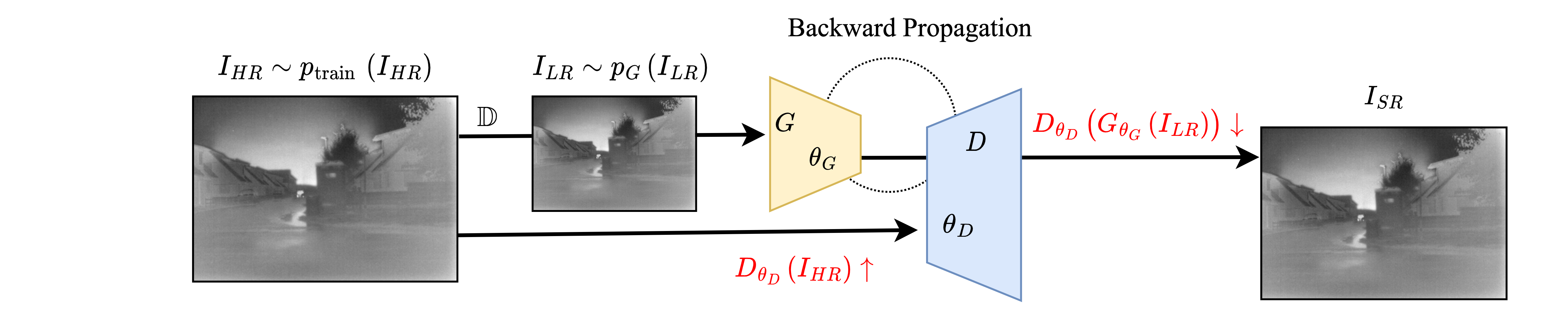}}
\caption{The adversarial training framework for IR image super-resolution. }
\label{fig3}
\end{figure}

The visualization of the adversarial generation framework for infrared images is shown in Fig.\ref{fig3}. Typically, we feed the generator $G$ with the low-resolution image $I_{LR}$, which is the high-resolution sample $I_{HR}$ after downsampling $\mathbb{D}$. Furthermore, the discriminator $D$ optimizes the objective function to achieve Nash equilibrium with the $G$ and finally generate the super-resolution image $I_{SR}$.


\section{Generative Adversarial Network and IR image SR\label{sec.3}}

After GAN was proposed, research based on generative models began to emerge in the field of super-resolution. Including SRGAN\cite{ledig2017photo}, ESRGAN\cite{wang2018esrgan}, and various other types of GAN models, they promote the development of this field together through the improvement of the model and the modification of mathematical analysis (WGAN\cite{gulrajani2017improved}). In the field of IR super-resolution, applications and research based on GAN models are also beginning to appear.

Before this, people have tried to directly use GAN models for normal images in infrared image reconstruction. Researchers  directly used SRGAN to reconstruct infrared images and obtained acceptable reconstruction images\cite{shao2018single}. However, the blurring of the edges and the lack of clarity of the details are still unacceptable.  Uses an improved DCGAN\cite{guei2018deep} to reconstruct infrared images. This work is different from directly using SRGAN to reconstruct images: first, it targets the reconstruction of face infrared images and proposes a self-built dataset. Secondly, the reconstruction effect is not compared with similar GAN methods, so it is difficult to explain the actual effect. The work of  compares the use of SR algorithms on IR images. Including SRGAN, ESRGAN, LapSRN\cite{lai2017deep}, RCAN\cite{zhang2018image}, and SRFBN\cite{li2019feedback}. The experimental results show that the SRFBN model has the best generalization ability, and for GAN models, there are always unpleasant artifacts due to inherent pattern defects. The GAN model used in normal images with more features has worse edge reconstruction effects in infrared images with fewer patterns. These works all show that the direct application of normal image algorithms in IR image reconstruction may encounter domain shift difficulties.

In the following sections, we describe the generative model in IR image super-resolution methods by module improvement (Sec.\ref{sec.3.1}) and introducing extra information (Sec.\ref{sec.3.2}), respectively.

\subsection{Module Improvement\label{sec.3.1}}

For module improvement, researchers focus on new modules and novel loss functions. In their work, \textit{Rivadeneira et al.\cite{rivadeneira2022novel}} propose attention modules and new loss functions. This framework based on Cycle-GAN was proposed and ResNet was introduced as a module for encoder. The attention module is used after the encoder. They further proposed to use Sobel edge detector as a new loss function to evaluate the similarity between images and to lead the network reconstruction.

Further, more modules related to attention mechanism are studied and proposed. The classical SeNet\cite{hu2018squeeze} network is used in IR image super-resolution tasks. Considering the network convergence difficulties, WGAN was introduced which is expected to overcome problems, by the gradient punishment\cite{huang2021pricai,liu2021infrared}. It is remarkable that the difference between the pixel attention mechanism and the channel attention mechanism will be further shown in the SR image quality.

\subsection{Introducing Extra Information\label{sec.3.2}}

After the model improvement, the community started to focus on introducing extra information for the IR image super-resolution task. Because IR images have limited feature information compared to visible images, introducing information from other sources will bridge the gap. For introducing information, the two categories mainly include: hybrid and separated. More details about these two different types of patterns will be discussed below. For the hybrid model, feature information from different sources is not purposefully distinguished. These features are coupled together in the feature space defined by the model parameters.

For example, researchers propose a multimodal visual thermal fusion model that aims to introduce high-frequency information from visible images to help reconstruct thermal images\cite{almasri2018multimodal}. Experiments have shown that this method using high-frequency information from visible images to reconstruct thermal images is helpful to improve the thermal image quality. There have been improvements in both objective assessment and subjective evaluation.

\begin{figure}[h]
\centerline{\includegraphics[width=\columnwidth]{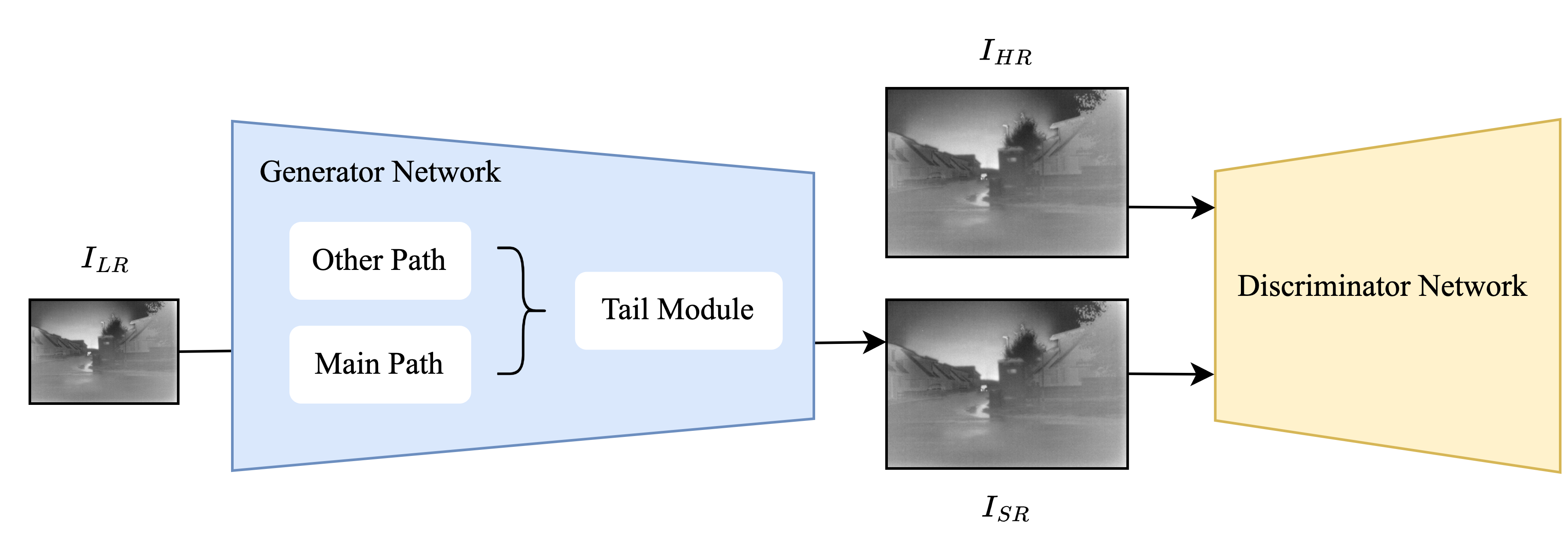}}
\caption{Transfer learning strategies are used in this framework to fuse the latent space between two domains, the visible image and the infrared image\cite{huang2021infrared}.}
\label{fig5}
\end{figure}

Moreover, the split model considering that different domain information should be employed more reasonably and the split type approach is further investigated.\cite{huang2021infrared} suggests a strategy using transfer learning to represent the specific patterns in the visible and infrared images in different modules (see Fig.\ref{fig5}). This information is fused in the tail module finally. This strategy illustrates the effectiveness from the view of mathematical analysis for splitting strategies. The experimental results also demonstrate competitive results.

There are some other types of approaches that also help to enhance the image quality. \cite{liu2019infrared} recommends that the image gradient should be more studied with its first layer features map extracted from VGG is used as the backbone as gradient information and further calculated loss function to guide the network to reconstruct the IR image. Other models recommend that the SR task should be considered with references to other tasks, such as denoising and detection\cite{batchuluun2020deep,mostofa2020joint}. For denoising, the framework proposed by Batchuluun is used for the detection task, but the network layers are deepened in the generator to achieve denoising and super-resolution. The detection framework suggested by \cite{mostofa2020joint} is also concerned with the correlation between multiple tasks. This work also uses a combined loss function to help enhance the image quality.

\section{Future Trends\label{sec.4}}

In this section, we will discuss the future trends in IR image super-resolution with GAN. First, new network architectures are helpful for improving the ability to represent patterns. Then, we will present promising directions: image quality assessment. Finally, unsupervised super-resolution in IR images will be discussed.

\subsection{Network Design\label{sec.4.1}}

With the explosive growth of deep learning, more and more new frameworks and models are being proposed and followed. Representative works include the \hys{Transformer} and diffusion models. 

For Transformers, its excellent performance in obtaining long-distance dependency information has made it widely used in various vision tasks and NLP tasks\cite{wei2022contrastive,yang2022focal,yu2022coca,ke2021musiq,herrmann2022pyramid,lee2021vitgan}. The results show that the introduction of Transformers into SR tasks has improved both subjective and objective indicators. It should be noted that in the IR image super-resolution task, we focus on the unique patterns of infrared images, such as gradient changes, which will bring new challenges to Transformers. In addition, the data-hungry cloud always hovers over the Transformer. Considering the difficulty of collecting IR image samples, we need to be more careful when introducing difficulties when introducing it into the IR super-resolution field. 

On the other hand, the good interpretability of the diffusion model makes it more attractive\cite{huang2021infrared,huang2021pricai,wang2018esrgan}. More importantly, it is also a generative model. The ability to fit the data distribution well attracts researchers to use it as a backbone network to build a neural network for IR image super-resolution.  More importantly, it is also a generative model. The ability to fit data distributions well attracts researchers to use it as a backbone network to build neural networks for IR image super-resolution.


\subsection{Unsupervised Image Super-Resolution\label{sec.4.2}}


Blind super-resolution, which involves reconstructing an HR image from a single LR image, has long been neglected in the field of super-resolution\cite{zhang2022closer,li2022face,yang2022degradation,zhou2022joint}. However, research on blind super-resolution has the potential to address the challenge of real-world super-resolution, where training samples for neural networks are often paired, unlike the degraded LR images that are commonly obtained from imaging devices, particularly in the case of infrared images.

Current approaches to solving the blind super-resolution problem focus on modeling the data distribution as accurately as possible in order to synthesize training data, which is a valuable approach\cite{lee2022learning,huang2022rethinking,zhang2021designing,wang2021real,son2021toward}. However, researchers could also consider improving the model by defining the structure of the generator and conducting further mathematical analysis. The interpretability of these types of methods could be beneficial for the continued growth and advancement of the field.

\section{Conclusion\label{sec.5}}


In this chapter, we discuss the potential applications and future directions of generative models in the context of IR image super-resolution. The use of generative adversarial networks (GANs) in super-resolution tasks has gained significant attention, particularly in the domain of IR images, which exhibit unique patterns. Researchers have made significant progress in this area through the development of novel module designs and the incorporation of additional information. It is anticipated that new generative models, such as diffusion models, will continue to drive advancements in this field.

\bibliographystyle{spmpsci.bst} 
\bibliography{ref}

\end{document}